\documentclass[conference]{IEEEtran}
\IEEEoverridecommandlockouts
\usepackage{cite}
\usepackage{amsmath,amssymb,amsfonts}
\usepackage{algorithmic}
\usepackage{graphicx}
\usepackage{textcomp}
\usepackage{xcolor}
\def\BibTeX{{\rm B\kern-.05em{\sc i\kern-.025em b}\kern-.08em
    T\kern-.1667em\lower.7ex\hbox{E}\kern-.125emX}}
\begin{document}


%

\newcommand*{\Scale}[2][4]{\scalebox{#1}{$#2$}}%
\newcommand*{\Resize}[2]{\resizebox{#1}{!}{$#2$}}%
\newcommand{\BlueText}[1]{\textcolor{blue}{#1}}
\newcommand{\remove}[1]{ }

\makeatletter
\setlength{\@fptop}{0pt}
\makeatother

\newcommand{\claudio}[1]{\textcolor{blue}{{#1}}}
\newcommand{\henry}[1]{\textcolor{red}{{#1}}}

\newcommand{\R}{\mathbb{R}}
\newcommand{\Z}{\mathbb{Z}}

\title{Data Architecture for Digital Object Space Management Service (DOSM) using DAT}

\author{
\IEEEauthorblockN{, {Moamin B. Abughazala}\thanks{moamin.abughazala@graduate.univaq.it}\IEEEauthorrefmark{2}, {Henry Muccini}\thanks{henry.muccini@univaq.it}\IEEEauthorrefmark{2},}

\IEEEauthorblockA{\IEEEauthorrefmark{2}\textit{DISIM Department},
\textit{University of L'Aquila},
L'Aquila, Italy}
}

\maketitle

\begin{abstract}
The Internet of Things (IoT) data and social media data are two of the fastest-growing data segments. Having high-quality data is crucial for making informed business decisions. The strategic process of leveraging insights from data is known as data-driven decision-making. To achieve this, it is necessary to collect, store, analyze, and protect data in the best ways possible. Data architecture is a complex task that involves describing the flow of data from its source to its destination and creating a blueprint for managing the data to meet business needs for information. In this paper, we utilize the Data Architecture Tool (DAT) to model data for Digital Space Management Service, which was developed as part of the VASARI project. This work focuses on describing the movement of data, data formats, data location, data processing (batch or real-time), data storage technologies, and main operations on the data.
\end{abstract}


\section{Introduction}
Italy is one of the famous country in the presence of rich cultural heritage sites, attracting millions of visitors every year. As per the latest data from Statista, the cultural heritage sites in Italy have attracted around 67 million visitors in the last two years \cite{Statista}. 
The visitors are also showing a growing interest to play an active and participatory role in the tourism experience, integrating the cultural content of the visit with self-generated personal content and sharing them with the community.

The International Data Corporation (IDC) \cite{idc} expects that by 2025 there will be more than 175 zettabytes of useful data.

 {\em Data Architecture} (DA) is an integrated set of specification artifacts used to define data requirements, guide integration, control data assets, and align data investments with business strategy. It also includes an integrated collection of master blueprints at different levels of abstraction \cite{10.5555/3165209}.

Cultural heritage in Italy, especially museums, is often under-exploited and does not provide adequate support for visitors in i) locating contents of interest; ii) discovering information on specific contents; iii) navigating within the heritage sites and iv) creating customized visit routes based on interests. Thereby limiting the affecting the overall experience of the visitor. To meet these needs, the VASARI project proposes a novel paradigm for sustaining onsite, immersive, inclusive, and contextualized user experiences
by exploiting several recent technologies, such as IoT and
mobile computing, semantics, big data processing, virtual and
augmented reality within an innovative infrastructure based
on microservices deployed in edge, fog and cloud nodes.

In this work, we present a DAT \cite{10.1007/978-3-031-36889-9_8} \cite{abughazala2023modeling} (Data Architecture Tool) to describe Digital Object Space Management Service (DOSM)  developed as a part of the VASARI project, DOSM provides support for the generation of a digital twin that captures the physical space and the cultural
assets, and it leverages IoT devices to enable navigation and localization support for visitors, thereby providing
them with an immersive visiting experience. DAT is for modeling data pipeline, data warehouses, big data architectures (Lambda and Kappa). That makes managing the visitors and cultural assets data more easily.

The paper is organized as follows. Background is presented in Section II. Our DAT architectural modeling methodologies are presented in Section III. The method is applied to a real case study in Section IV, and the conclusions are finally drawn in Section V.

\section{Background}

CAPS is an architecture-driven modeling tool for IoT. CAPS  aims to support the architecture description, reasoning, design decision process, and evaluation of the CAPS architecture in terms of data traffic load, battery level, and energy consumption of its nodes.  This environment is composed of two important parts, First, the CAPS modeling framework \cite{muccini2017caps}, and second, the CAPS code generation framework \cite{sharaf2017architecture} \cite{sharaf2018arduino} \cite{sharaf2017simulating}. The CAPS modeling framework includes software architecture modeling language (SAML), hardware modeling language (HWML), and physical space modeling language (SPML).

DAT \cite{10.1007/978-3-031-36889-9_8} is an architecture modeling frameworks for describing the data flow between data nodes. It has the ability to describe two level of architectures {\em High Level Architecture (HLA)} and {\em Low Level Architecture (LLA)}. It has the abilities to describe different cases, for example (Data Pipeline, Data Warehouses, Big Data architectures like Lambda and Kappa). It is considered as a data view for CAPS. 



\section{Methodology}

This section describe the data architecture of IoT applications through the Data Modeling Language (DAML).
By defining the underlying meta-model for DAML, then we can formalize the structure and constructs of the DAML language. 

Any {\tt DataArchitecture} of IoT can contain a set of {\tt DataNodes} (components) and {\tt Connections}. A Component is considered a computational unit that has an internal state and a known interface. The internal data state of a component is denoted by the current behavioral of  data representation and its values. A {\tt DataRepresentation} can be described as a list of internal data elements; data representation is represented by actions and events defined in the component behavior, such as {\tt SendData}, {\tt ReceiveData}, etc.
 
 Data representation includes data formats, data storage technologies, location and processing type.
   Our meta-model covers all {\tt DataFormats} to describe data diversity of different data sources, {\tt Structured} ({\tt RelationalDB}) , {\tt Semi-Structured}  ({\tt Email, SMS, CSV, JSON, XML} and etc.) and {\tt Unstructured} ( {\tt GPS data}, {\tt Multi-media} and {\tt Office Files}).
   Data {\tt ProcessingType} to describe how data is going to be processed as a {\tt Batch} or {\tt Real-time}. To describe where data will be stored we use data storage technologies {\tt DataStorageTech.} that includes {\tt NoSQL} Databases ({\tt Document, Key-value, graph and column}), {\tt NewSQL} Databases ({\tt Historical, Real-Time, Stream, Timestamp}), and {\tt File System}  ({\tt HDF, GFS, AFS, GPFS and Blobseer}).
   Moreover, {\tt Location} is used to describe the location of data node on cloud or locally.
   A {\tt NodeBehavior} indicates the current status of the component that describes the data in a specific data node. For example in a generation data node, we can find elements that describe the source of the data and the format. Every NodeBehavior has a set of behavioral elements denoted by actions, and events that all together depict the data flow within the component.
   
   Data nodes can interact by passing data through data ports ({\tt DataPort}).  For receiving incoming data input data ports ({\tt InDataPort}) are used while output data ports ({\tt OutDataPort}) are used for sending outgoing data. 
   In this context, a connection represents a unidirectional communication channel between two data ports of two different components.
  An action is considered an important behavioral element that represents an atomic task that could be performed inside the data node. This element can be executed when a previous action in the behavioral data flow has been achieved or it could be triggered by an event like {\tt ReceiveData}. 
  
  The main actions of data behavioural elements to describe a data behavior inside the data node are as follows, Generation to represent the source of the data. Ingestion describe how data can move from source to data lake. Process include a list of sub-processes that could be used to describe a complete processing node. Store shows the main tasks to save, retrieve, archive and govern the data. Analyze, used to describe which type or technique could be used for analyzing the data. Consume to show how data could be consumed  like visualize, reports and API.
  
  An {\tt Event} is triggered in response to the external
stimulus of the component. To show the data flow and connection between the events and actions , we use links, and we could use them to decide the order in which actions can be performed and which one must be executed directly after an event.

\section{Results}
\begin{center}

\begin{figure}[htbp]
	\centering
 \caption{The Data architecture for DOSM}
 \makebox[\textwidth]
	{
	   \includegraphics[scale=0.26]{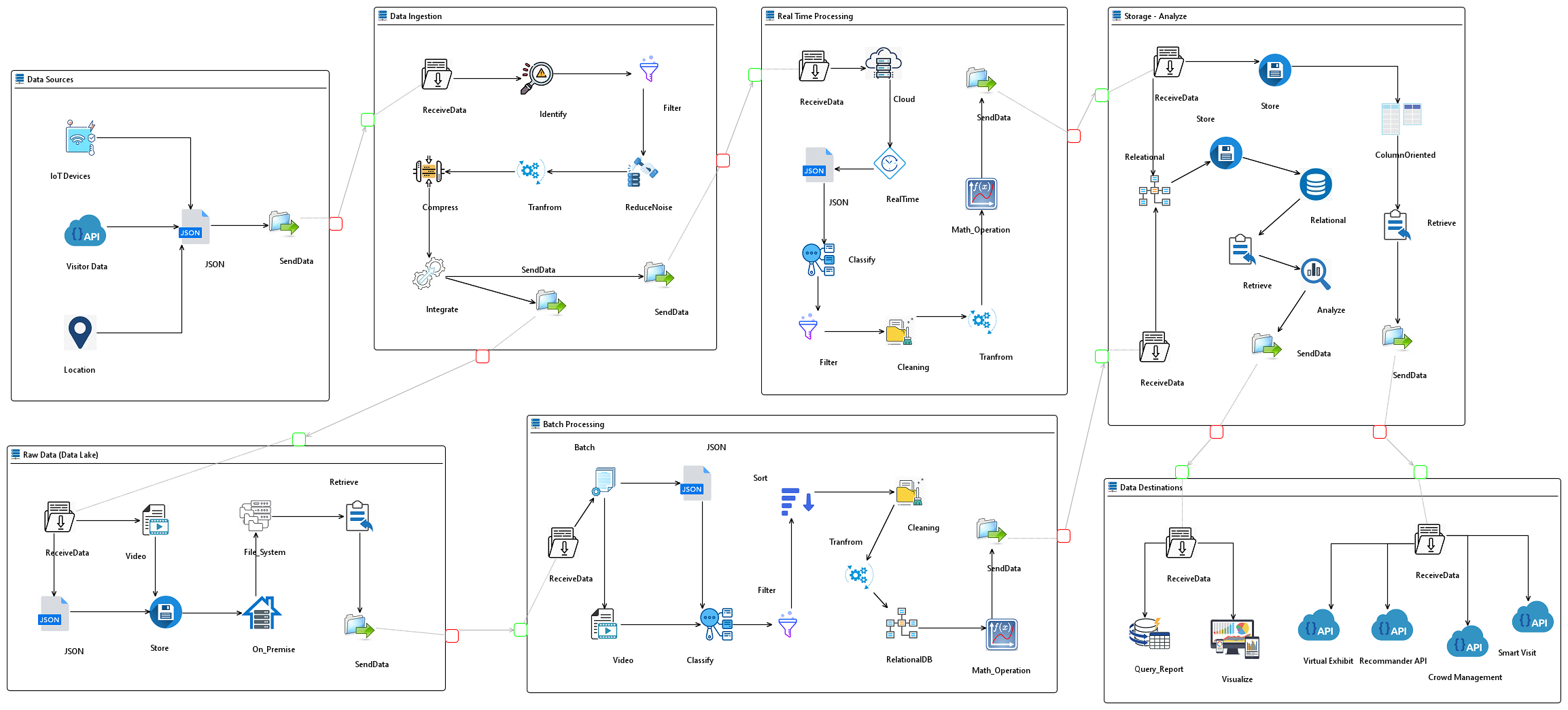}
    }
	
	\label{fig:vasari}
\end{figure}
    
\end{center}

This section shows the DAT of VASARI project case study. From a structural point of view, Figure \ref{fig:vasari}  shows the main six data nodes; Data sources, Data Ingestion, Raw Data, Real-time processing, Batch Processing, Storage and Visualize Data.

The \textit{Data Sources} node is responsible for representing the rule of generating data from different data sources (IoT devices, User's Mobile data) and the data format from these sources (JSON) and transferring the data to \textit{Data Ingestion} node. It describe how data extracted from various data sources and loaded into downstream (responsible of making the data available to the data pipeline), which includes identify, validate, compress, reduce the noise, and transform into a uniform data format. After that, the data will move to two data nodes Raw Data and Real-Time Processing. For \textit{Raw Data} node, it is responsible about saving the data in the natives formats locally using one of File System storage technologies like (HDFS) to be used as a historical data and to fed the batch Processing node. 

For \textit{Real-Time} and \textit{Batch Processing}, they have similar processing functionalities including classify the incoming data, filter, clean, and transform to another format to be stored in different kind of DBs to be compatible for later usage. The real-time processing is working whenever the data is available to be able to handle continuous stream data. Whereas, Batch Processing wait a specific amount of time to handle the collected group of data. After that, in the \textit{Storage and Analyze} phase, the data will be stored in different kind of databases (Relational and Column Oriented) to be ready to serve any query or any other data consumption or to be ready for any kind of analysis. 

Finally, The data will be available for visualization and other systems, Recommendation, Crowd Management, Smart Visiting. 

\section{Conclusion}
In this work, we presented Data Architecture Tool to describe Digital Object Space Management Service (DOSM) developed as a part of the VASARI project.
\bibliographystyle{unsrt}  
\bibliography{bib}

\end{document}